\documentclass{aa}  
\usepackage{natbib}
\usepackage{graphicx}
\usepackage{gensymb}
\usepackage[toc,page]{appendix}
\usepackage{pbox}
\usepackage{makecell}
\usepackage[flushleft]{threeparttable}
\usepackage{hyperref}
\usepackage{multirow}
\usepackage{multicol}
\usepackage{subcaption}
\hypersetup{colorlinks=true,breaklinks=true,citecolor=blue}
\usepackage{xcolor}
\usepackage{lscape}
\usepackage[export]{adjustbox} 
\usepackage{txfonts}
\bibpunct{(}{)}{;}{a}{}{,}

\newcommand{\xmm}{XMM-\textit{Newton} }
\newcommand{\nustar}{\textit{NuSTAR} } 
\newcommand{\chandra}{\textit{Chandra} }
\newcommand{\apm}{APM 08279 }
\newcommand{\xstar}{\textit{Xstar} }
\newcommand{\xspec}{\textit{Xspec} }

\defcitealias{lanzuisi2019}{L19}
\defcitealias{chartas2009}{C09}
\defcitealias{hasinger2002}{H02}

\begin{document} 

\title{
	The properties of the X-ray corona in the distant ($z=3.91$) quasar APM 08279+5255
	}
\titlerunning{APM 08279+5255}

   \author{E. Bertola\inst{1}\fnmsep\inst{2}\thanks{elena.bertola2@unibo.it}
             \and
          C. Vignali\inst{1}\fnmsep\inst{2}
             \and
          G. Lanzuisi\inst{2}
          \and
          M. Dadina\inst{2}
          \and
          M. Cappi\inst{2}
          \and
          R. Gilli\inst{2}
          \and
          G. A. Matzeu\inst{1}\fnmsep\inst{2}\fnmsep\inst{3}
          \and
          G. Chartas\inst{4}
          \and
          E. Piconcelli\inst{5}
          \and
          A. Comastri\inst{2}
          }

   \institute{Dipartimento di Fisica e Astronomia ``Augusto Righi'', Universit\`a degli Studi di Bologna, via P. Gobetti 93/2, 40129 Bologna, Italy
         \and
             INAF--OAS, Osservatorio di Astrofisica e Scienza dello Spazio di Bologna, via P. Gobetti 93/3, 40129 Bologna, Italy
          \and
        European Space Agency (ESA), European Space Astronomy Centre (ESAC), E-28691 Villanueva de la Ca\~{n}ada, Madrid, Spain
        \and 
            Department of Physics and Astronomy of the College of Charleston, Charleston, SC 29424, USA
        \and
            INAF – Osservatorio Astronomico di Roma, Via Frascati 33, 00040 Monte Porzio Catone, Roma, Italy       
             }

   \date{Received ; accepted }
 
 \abstract
  {
  We present new joint \xmm and \nustar observations of APM 08279+5255, a gravitationally-lensed, broad-absorption line quasar ($z=3.91$). After showing a fairly stable flux ($f_{\rm2-10}\simeq4-5.5\times10^{-13}\rm~erg~s^{-1}$) from 2000 to 2008, \apm was  found in a fainter state in the latest X-ray exposures ($f_{\rm2-10}\simeq2.7\times10^{-13}\rm~erg~s^{-1}$), which can likely be ascribed to a lower X-ray activity. Moreover, the 2019 data present a prominent Fe K$\alpha$ emission line and do not show any significant absorption line. This fainter state, coupled to the first hard X-ray sampling of APM 08279+5255, allowed us to measure X-ray reflection and the high-energy cutoff in this source for the first time. From the analysis of previous \xmm and \chandra observations, X-ray reflection is demonstrated to be a long-lasting feature of this source, but less prominent prior to 2008, possibly due to a stronger primary emission. 
  The estimated high-energy cutoff ($E_{\rm cut}=99_{-35}^{+91}$~keV) sets a new redshift record for the farthest ever measured and places APM 08279+5255 in the allowed region of the compactness-temperature diagram of X-ray coronae, in agreement with previous results on high-$z$ quasars. 
  }


  \maketitle
   
%

\section{Introduction}
Observational efforts in the past three decades have demonstrated the validity of the two-phase model \citep{haardt1991,haardt1994} in describing the high-energy emission of Active Galactic Nuclei (AGN). Optical/UV disc photons are predicted to be Compton up-scattered by the electrons of the hot corona ($T_e\sim10^{8-9}K$), which surrounds the central supermassive black hole (SMBH). The "Comptonization" process generates the cutoff-power-law-like spectrum measured in the X-rays, in which the cutoff energy is set by the temperature of the hot corona. At very hard energies, photon-photon collisions decay into electron-positron pairs which can, in turn, annihilate and produce energetic photons. 
Pair production can then become a runaway process acting as a natural thermostat for the corona. The conditions in which this takes place depend on a combination of corona temperature and radiative compactness \citep{cavaliere1980}, as well as on the plasma optical depth ($\tau$). In fact, Comptonization models of hot coronae with slab geometry predict a cutoff in the X-ray power law at $E_{\rm cut}/k_{\rm B}T_{\rm e}\simeq2(3)$ for optically thin(thick) plasma \citep[e.g.,][]{petrucci2001}.
The typically employed quantities \citep{cavaliere1980,guilbert1983} include the dimensionless temperature parameter:
\begin{equation}
    \theta=\frac{k_{\rm B}T_e}{m_ec^2}=\frac{E_{\rm cut}}{Km_ec^2}
    \label{eq:theta}
,\end{equation}
with $K=2(3)$ for an optically thin(thick) plasma, along with the dimensionless compactness parameter: 
\begin{equation}
    \ell=\frac{L_{\rm X}}{R_{\rm X}}\frac{\sigma_T}{m_ec^3}
    \label{eq:compactness}
,\end{equation}
where $m_{\rm e}$ and $T_{\rm e}$ are electron mass and temperature, respectively, $L_{\rm X}$ and $R_{\rm X}$ are the luminosity and size of the X-ray source, respectively, $\sigma_{\rm T}$ is the Thomson scattering cross-section, and $k_{\rm B}$ is the Boltzmann constant. 

\renewcommand{\arraystretch}{1.15}
\begin{table*}[!h]
        \begin{center}
                \caption{Log of APM 08279+5255 observations from 2019}
                \label{tab:red}
                \begin{tabular}{ccccc}
                        \hline\hline
                        Observation & ObsID             & Date        & Net Exposure (ks)   & $f_{2-10}$                      \\ \hline
                        XMM 101     & 0830480101        & 2019 \textit{Mar} 24 &  24.1      & $2.9_{-0.3}^{+0.1}$                                                      \\
                        Nu02      & 60401017002         & 2019 \textit{Apr} 19 &  93.5 | 92.8       & $2.8_{-0.4}^{+0.2}\, |\, 3.2_{-0.4}^{+0.3}$                          \\
                        Nu04      & 60401017004         & 2019 \textit{Apr} 22 &  59.7 | 59.2       & $ 2.6_{-0.4}^{+0.4}\, |\, 3.4_{-0.6}^{+0.3}$                         \\ 
                        XMM 301     & 0830480301        & 2019 \textit{Apr} 23 &  24.5 | 28.2 | 25.7 & $2.3_{-0.2}^{+0.1}\, |\, 2.4_{-0.3}^{+0.1}\, |\, 2.5_{-0.3}^{+0.2}$  \\ \hline
                \end{tabular}
        \end{center}
        \tablefoot{Values of XMM 101 refer to EPIC-pn only; values of XMM 301 refer to EPIC-pn, -M1, -M2 respectively; values of \nustar refer to FPMA, FPMB, respectively. Observed-band 2--10 keV absorbed flux ($10^{-13}$\,erg\,cm$^{-2}$~s$^{-1}$, errors at 90\% confidence level) is estimated using {\sf acutpl} model. Net exposure: exposure time after cleaning the event file from flare events.}
\end{table*}    
\renewcommand{\arraystretch}{1.0} 

In principle, by populating the $\ell-\theta$ diagram, it is possible to probe the mechanisms regulating the corona temperature and test the pair-production thermostat predictions. To that aim, broadband X-ray spectra of compact sources are needed to properly model the primary emission and, in particular, its high-energy cutoff. This was extensively done with hard X-ray observatories, such as \textit{BeppoSAX} \citep[e.g.,][]{petrucci2001,dadina2007,dadina2008}, \textit{INTEGRAL} \citep[e.g.,][]{malizia2014}, and \textit{Swift} \citep[e.g.,][]{vasudevan2013,koss2017}. A real breakthrough in the study of X-ray coronae has arrived thanks to \textit{NuSTAR} \citep{harrison2013}, the first focusing hard X-ray telescope, which allowed for an improved estimation of the coronal temperature in nearby sources. By gathering literature results from both non-focusing instruments and \textit{NuSTAR}, \citet{fabian2015} built a compilation of $E_{\rm cut}$ measurements in both local AGN and black-hole binaries, finding many of their coronae to lie at the edge of the runaway pair-production region in the $\ell-\theta$ plane. Similar results were later obtained by \citet{ricci2018} based on sources of the \textit{Swift}/BAT AGN Spectroscopic Survey (BASS, \citealp{ricci2017}). Interestingly, \citet{ricci2018} also discovered a negative correlation between the average high-energy cutoff and Eddington ratio in BASS AGN, regardless of either luminosity or SMBH-mass selection. 

With its wide hard X-ray bandpass, \textit{INTEGRAL} allowed for detailed studies of the coronal high-energy cutoff in local AGN \citep[e.g.,][]{molina2009,molina2013,deRosa2012}, many of which were later updated and confirmed by \nustar measurements. Nonetheless, \nustar can properly constrain the high-energy cutoff only if this falls in its bandpass or, otherwise, only in high count-rate sources. The outstanding results obtained by \nustar have thus been restricted to nearby AGN ($z\lesssim0.1$) with $E_{\rm cut}\lesssim200\rm~keV$ and $L_{\rm X}\lesssim10^{45}~\rm erg~s^{-1}$ \citep[see, e.g., the recent work by][]{akylas2021}, until recently, when \citet{lanzuisi2019} (hereafter, \citetalias{lanzuisi2019}) were able to probe the high-luminosity regime ($L_{\rm X}>2\times10^{45}~\rm erg~s^{-1}$) through \nustar observations of high-$z$ AGN. 
\citetalias{lanzuisi2019} were the the first to proper constrain the high-energy cutoff of two AGN at $z\gtrsim2$ (2MASSJ1614346+470420 at $z=1.86$ -- hereafter, 2MASSJ16; B1422+231 at $z=3.62$ -- hereafter B1422) using \nustar data, confirming the measurement for B1422 by \citet{dadina2016} from \xmm data. 
Both sources show rather low $E_{\rm cut}$ values ($\lesssim100\rm~keV$) and fall in the limited allowed region for high-luminosity AGN of the $\ell-\theta$ plane. Interestingly, the measured $E_{\rm cut}$ values are much lower than those of BASS AGN showing similar Eddington ratios ($E_{\rm cut}\sim150-170\rm~keV$, \citealt{ricci2018}). 

One of the most interesting high-$z$ quasars is \object{APM 08279+5255} ($z=3.91$, APM 08279 hereafter; \citealt{irwin1989}). This broad-absorption line quasar is lensed in three images by an as-yet-undetected foreground galaxy, possibly set at $z=1.06$ \citep{ellison2004}. Due to the lensing-system uncertainty, different models predict very different magnification values, ranging from $\mu_{\rm L}=4$ \citep{riechers2009} to $\mu_{\rm L}=100$ \citep{egami2000}. Regardless of the actual magnification factor, \apm is among the brightest high-$z$ AGN in many bands, with one of the best sampled high-$z$ spectral energy distributions \citep[e.g.,][]{stacey_2018,leung2019}. In fact, \apm is a very well known quasar in many astrophysical research fields and it lately became the first high-$z$ quasar whose SMBH mass was estimated via reverberation mapping of the \ion{Si}{IV} and \ion{C}{IV} emission lines ($\log(M_{\rm BH}/M_{\odot})=10\pm0.1$, \citealp{saturni2016}). 
\apm is also a very peculiar source for ultra-fast outflows (UFOs), namely, the X-ray winds that could be responsible for the generation of galaxy-wide outflows and thus for the establishment of the AGN-host-galaxy co-evolution \citep[e.g.,][]{faucher-giguere2012,kingpounds2015,costa2020}. 
In fact, \apm was the first high-$z$ source in which UFOs were detected \citep{chartas2002} and it was later found to host  some of the fastest X-ray winds ever seen ($v_{\rm out}$ up to 0.76\textit{c}, \citealt{chartas2009}, hereafter \citetalias{chartas2009}). However, the most remarkable feature of \apm is the double-velocity UFO present in all the observations up to early 2008 \citep[\citetalias{chartas2009};][]{saez2009}, except for its first X-ray exposure (\citealt{hasinger2002}, hereafter \citetalias{hasinger2002}). 

We present in this paper the first X-ray broadband analysis of APM 08279, making use of the latest \xmm observations followed up by the first ever \nustar exposures of this source (2019, PI: G. Lanzuisi). The paper is organized as follows: reduction and analysis of 2019 observations are discussed in Sects. \ref{sec:data_red}--\ref{sec:2019_data}. Our results are then compared to previous \chandra and \xmm observations, which we re-analyzed, in Sect. \ref{sec:old_data}. We then place the observed X-ray corona properties in a broader context in Sect. \ref{sec:discussion}. The scientific results are summarized in Sect. \ref{sec:summary}. We assume a flat ${\Lambda}$CDM cosmology \citep{planckcoll2020}, with $H_{\rm 0}=70.0\ {\rm km\ s^{-1}\ Mpc^{-1}}$ and $\Lambda_{\rm 0}=0.73$ throughout the paper.

\section{Data reduction}
\label{sec:data_red}
\apm was observed by \xmm on 2019 \textit{March} 24 for 31.4 ks (hereafter, XMM 101). On that date, EPIC-MOS cameras failed, thus it was observed again on 2019 \textit{April} 23 for additional 33.3 ks (hereafter, XMM 301). These exposures were followed up by \nustar on 2019 \textit{April} 19 for 93.5 ks (hereafter, Nu02) and on 2019 \textit{April} 22 for 59.7 ks (hereafter, Nu04). The observation log is shown in Table \ref{tab:red}. 

\xmm data were reduced applying standard procedure and the latest calibration files through SAS v.18.0. The event files of EPIC-pn cameras were filtered at 1.2 and 1.0 counts per second in the 10--12 keV band, for XMM 101 and 301, respectively, while those of XMM 301 EPIC-MOS cameras were filtered at 0.3 counts per second, in the same band. EPIC-pn source spectra were extracted from circular regions of 25\arcsec radii for both XMM 101 and 301 ($\simeq80$\% encircled energy fraction); EPIC-MOS source spectra were extracted using 20\arcsec-radius circles ($\simeq75$\% encircled energy fraction). Background spectra were extracted from circular regions of 60\arcsec radii for each \xmm camera. Wider source regions, coupled with different good-time-interval filtering thresholds and other background extraction regions, were tested. No significant improvement of the spectral signal-to-noise ratio (S/N) was yielded, therefore, we stuck to the filtering and spectra extraction setup just described (i.e., source regions encircling the PSF core). 
\begin{figure}[!h]
        \centering
        \includegraphics[width=0.98\linewidth
        ]{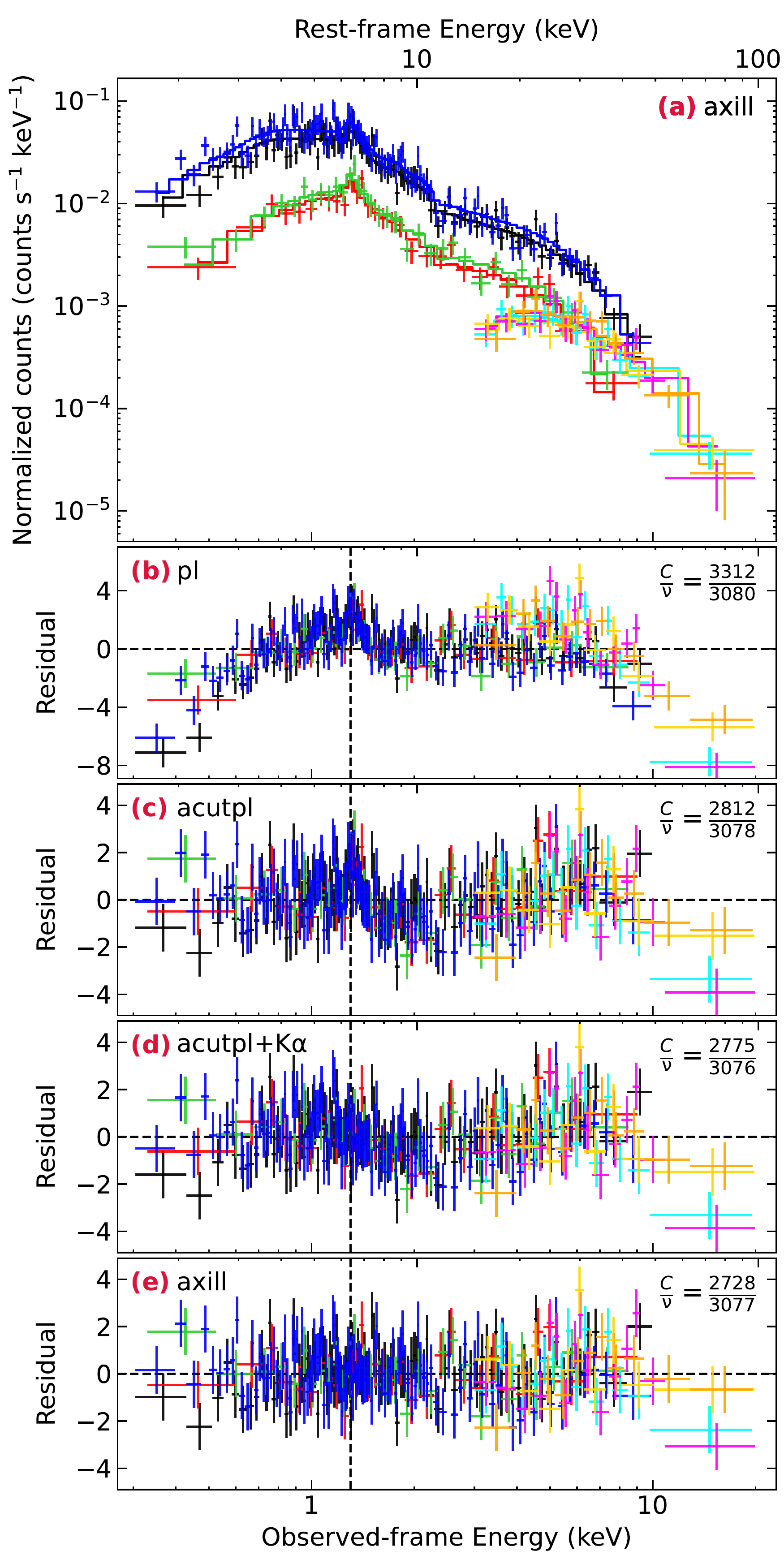}
        \caption{Broadband X-ray spectra of \apm collected by \xmm and \nustar in 2019. Panel (a): Observed-frame spectra and best-fit model ({\sf axill} model). Panels from (b) to (e): Observed-frame residuals. Vertical black dashed lines mark the energy of the Fe K$\alpha$ emission line. Spectra in each panel were rebinned to 4$\sigma$ (with {\sf setplot rebin 4 100} in {\tt Xspec}) for showing purposes. Data are color-coded as follows: XMM 301 EPIC-pn is shown in black, XMM 301 EPIC-MOS1 in red, XMM 301 EPIC-MOS2 in green, XMM 101 EPIC-pn in blue, Nu02 FPMA in cyan and FPMB in magenta, Nu04 FMPA in yellow, and FPMB in orange. Model from top to bottom: (a) {\sf axill} model, (b) {\sf pl} model, (c) {\sf acutpl} model, (d) {\sf acutpl+K$\alpha$} model, (e) {\sf axill} model. All models are modified by Galactic absorption ($N_{\rm H}=4.2\times10^{20}\rm\,cm^{-2}$). Best-fit parameters are summarized in Tables \ref{tab:models_recap}. } 
        \label{fig:residui}
\end{figure} 

\nustar observations were processed using the standard pipeline of \nustar Data Analysis Software package (NuSTARDAS) v.2.0.0 (within Heasoft v.6.28) and calibrated with \nustar CALDB v.20200813. No significant background flares are present in these observations -- a fact that we checked through the IDL script {\tt nustar\_filter\_lightcurve}\footnote{\url{https://github.com/NuSTAR/nustar-idl}}. After testing different extraction regions to find the ones yielding the best S/N, we selected 40\arcsec-radius circles for the source ($\simeq60$\% encircled energy fraction). These were coupled to annular background regions, centered on the target, with inner(outer) radii of 110\arcsec(170\arcsec) to exclude the wings of the source PSF and sample the non-uniform local background. This extraction setup was used for both FPMA and FPMB in each observation. Unfortunately, \apm turned out to be fainter than expected based on past observations (as described in Sect. \ref{sec:old_data}). Thus, during these exposures, source spectra are background dominated above $15-20$\,keV regardless of the reduction and spectra-extraction setup. 

\section{Spectral analysis of 2019 data}
\label{sec:2019_data}
Having checked that no significant intra-observation flux variability is present, we fit the time-averaged spectra with \xspec v.12.11.1. EPIC-pn data belong to the high-statistics regime, whereas EPIC-MOS and \nustar spectra to the mid-to-low-statistics regime  ($\simeq$600 and 300 net counts, respectively). 
We grouped our data to 1~count/bin and applied \textit{C}-statistics \citep{cash1979} because matching the requirements to use the $\chi^2$ statistics (at least 20~cts/bin) would have led to a loss in the energy resolution of \nustar and EPIC-MOS due to the coarse binning. 
Nevertheless, we also tested our models on spectra grouped at 20~cts/bin using $\chi^2$ statistics as a sanity check and found results consistent to those presented here. 

XMM 301 and the two \nustar observations are almost simultaneous, while XMM 101 was taken 30 days before XMM 301. No significant spectral variability is present between the two \xmm epochs\footnote{The ratio of the two EPIC-pn spectra shows no evident trend and is well consistent with being constant. }. Thus, we fit spectra from the four epochs together by linking all parameters, if not said otherwise, letting cross-calibration constants free to vary. 
Throughout the paper, all models are modified by Galactic absorption \citep[$N_{\rm H}=4.2\times10^{20}\rm\,cm^{-2}$, ][]{HI4PI2016_nh} and errors and upper limits are given at 90\% confidence level, unless otherwise stated. 

\subsection{Broadband X-ray spectra}
\label{sec:continuum}
The joint fit of \xmm and \nustar data allows us to model the broadband-band continuum of \apm in the 0.3--20 keV observed-frame energy range (i.e., $\sim$1.5--98 keV rest-frame energy range). 
The spectra present a cutoff at hard energies and soft absorption in excess of the Galactic value, which are clearly visible in the residuals against a single power-law model ({\sf pl} Model; see Fig. \ref{fig:residui}, panel (b)). In fact, a {\tt zphabs*zcutoffpl} model (hereafter, {\sf acutpl} model; Fig. \ref{fig:residui}, panel (c)) well reproduces both features in statistical terms, but yields a power law that is considerably harder ($\Gamma\simeq1.3$) than expected values for an AGN \citep[e.g.,][]{vignali1999,piconcelli2005,just2007} and a very low high-energy cutoff ($E_{\rm cut}\simeq34 \,$keV). Absorption in the soft band is due to a ``cold'' medium placed at the systemic redshift of the source, with a column density consistent with previous observations \citepalias{hasinger2002,chartas2009}. 
For the rest of the analysis, our models include this additional absorption component. The data clearly present a prominent Fe K$\alpha$ emission line, detected at $E_{\rm rest}=6.5\pm0.1$ keV ($E_{\rm obs}\simeq1.3$ keV) as a highly-significant ($\Delta Cstat/\Delta\nu=37/2$) narrow line\footnote{The line width is visually resolved, but letting this parameter free to vary yields no statistical improvement ($\Delta Cstat/\Delta\nu=1/1$). Moreover, the best-fit line width is consistent with being narrow both based on its face value ($\sigma=0.15$ keV) and its 90\% confidence range ($\sigma<0.7$ keV) in the rest frame. We thus set the rest-frame width of the Fe K$\alpha$ line to $\sigma=0.1$ keV.} on top of an {\sf acutpl} continuum in both \xmm observations. Therefore, we tie the Fe K$\alpha$ line component between the two epochs. 

\begin{figure}[!h]
        \centering
                \adjincludegraphics[clip,trim={{.1\width} 0 0 {.17\width}}
                ,width=0.75\linewidth, angle=-90
                ]{images/cont_ecut_ga_mod_xillzph_mos_frxi.pdf}
                \\
                \adjincludegraphics[clip,trim={{.1\width} 0 0 {.17\width}},width=0.75\linewidth, angle=-90
                ]{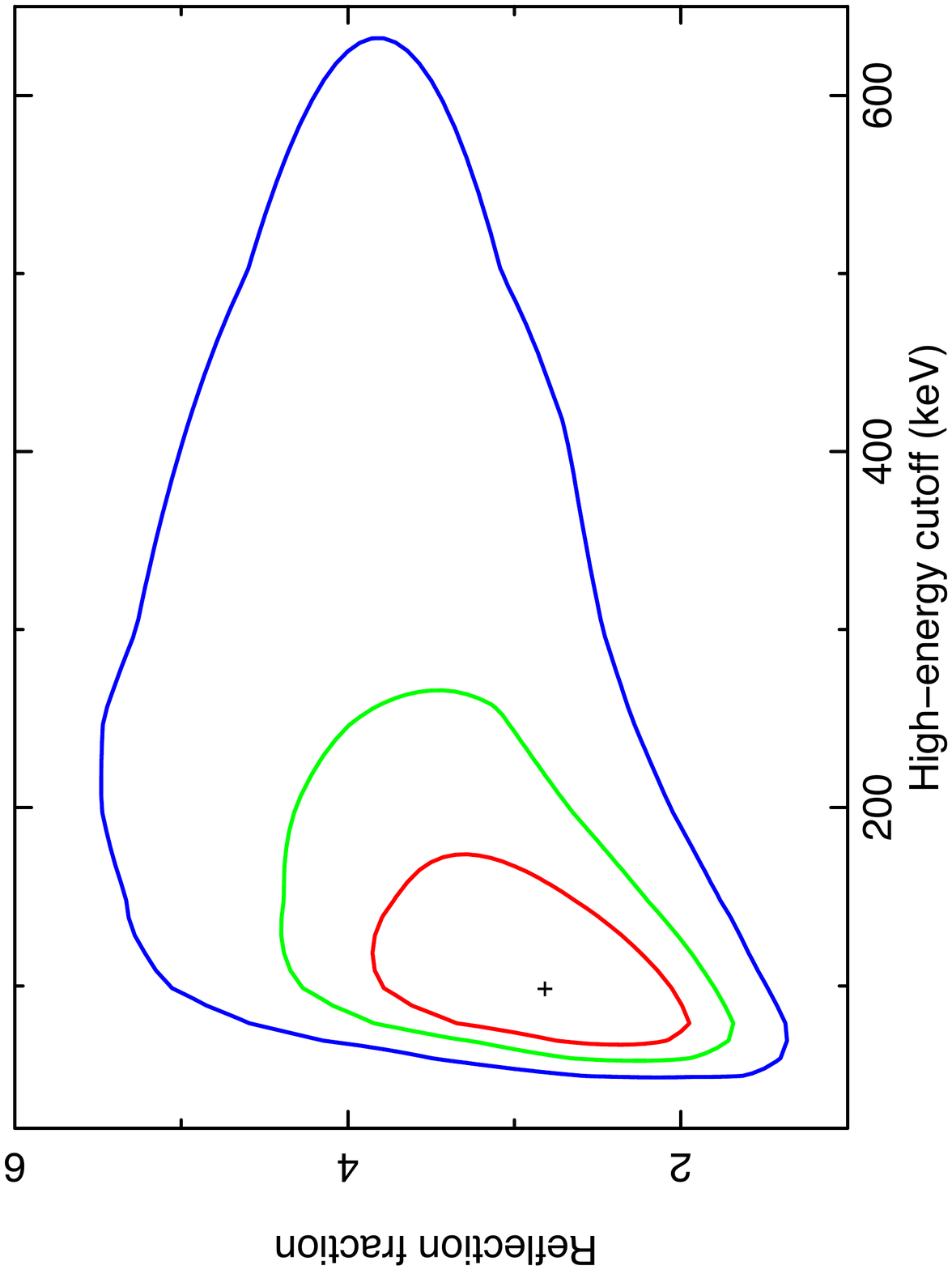}
        \caption{Confidence contours of reflection parameters obtained with an {\sf axill} model (see Sect. \ref{sec:continuum} and Table \ref{tab:models_recap}). \textit{Top:} High-energy cutoff vs. photon index. \textit{Bottom:} Reflection fraction vs. high-energy cutoff. Contours are color-coded as follows: red, green, blue for 68\%, 90\%, 99\% confidence level, respectively. } 
        \label{fig:refl_cont}
\end{figure}

An {\sf acutpl+K$\alpha$} model fails to fit the high-energy bump (Fig. \ref{fig:residui}, panel (d)) which, combined with the low photon index and high Fe K$\alpha$ equivalent width (EW; rest-frame $\rm EW=318^{+94}_{-90}$ eV) can be evidence for X-ray reflection. We tested this scenario through the non-relativistic reflection model {\tt xillver}, part of the {\tt relxill} package \citep{garcia2014,dauser2014}, which accounts for a direct cutoff power law and its reprocessed emission (continuum and self-consistent emission and absorption features) by a distant, (possibly ionized) medium.  
Reflection is parameterized through photon index $\Gamma$; high-energy cutoff, $E_{\rm cut}$; iron abundance (that we set to Solar); ionization of the disk, $\log\xi$\footnote{The ionization parameter $\xi$ is defined as  $\xi = L_{\rm X}/nr^2$, where $L_{\rm X}$ is the X-ray luminosity of the incident radiation,  $n$ is the gas density and $r$ is the distance from the ionizing source. }; inclination angle, \textit{i}; and reflection fraction, $R$, defined as the fraction relative to the reflected emission expected from a slab subtending a 2$\pi$ solid angle. Low inclination angles return a better fit and are naturally preferred by the data when the parameter is set free to vary; however, the data statistics prevents us from actually constraining it, thus we set $i=30$\textdegree. 
Similarly, we first leave the ionization of the disk free to vary and then freeze its value to its best fit ($\rm\log(\xi/erg~s^{-1}~cm )=1.7$). 
By the inclusion of the reflection continuum ({\sf axill} model), we find a better representation of our broadband spectra, both statistically ($\Delta Cstat=47$ for one additional parameter; see confidence contours in Fig. \ref{fig:refl_cont}) and physically (see Table \ref{tab:models_recap}). 
The power-law photon index ($\Gamma=2.1_{-0.2}^{+0.1}$) agrees with typical values of high-$z$ sources \citep[e.g., ][]{vignali2005,just2007} and so does the high-energy cutoff ($E_{\rm cut}=99_{-35}^{+91}$ keV) with the few other measurements available at $z>1$ (\citealt{lanzuisi2016}; \citealt{dadina2016}; \citetalias{lanzuisi2019}). 
The yielded reflection fraction (${R=2.8_{-0.9}^{+1.1}}$) carries the information that the reflecting material is seeing a source primary emission that is much larger than the one reaching the observer. Two of the possible explanations for such a large value of $R$ are \textit{i)} pc-scale reflection in which the primary source activity has dropped (and so has the direct emission seen by the observer), whereas the reflector is still illuminated by the echo of the previous stronger source emission due to the travel-time delay \citep[e.g.,][]{lanzuisi2016}; \textit{ii)} disc reflection in a lamp-post geometry where the corona height is low and thus light-bending is severe \citep[e.g.,][]{gandhi2007}. 

\renewcommand{\arraystretch}{1.15}
\begin{table*}[]
        \caption{Summary of the best-fit parameters of each model tested on 2019 data. }
        \centering
        \begin{tabular}{lccccccc}
                \hline\hline
                Model                 &   ${\rm \Gamma}$    &    $N{\rm_H}$     &  $E_{\rm cut}$   & $E{\rm _{Fe\,K\alpha}}$ & $EW{\rm _{Fe\,K\alpha}}$ &            $R$             & $C$stat ($\nu$) \\
                                 (1)    &   (2)     &  (3)   & (4) & (5) &          (6)           & (7) & (8) \\ \hline
                {\sf pl}              & ${\rm 1.35\pm0.03}$ &        --         &        --        &           --            &            --            &            --             &  3312 (3080)   \\
                {\sf acutpl}           &  ${\rm 1.3\pm0.1}$  & ${\rm4.7\pm0.8}$  & $36_{-7}^{+10}$  &           --            &            --            &            --             &  2812  (3078)  \\
                {\sf acutpl+K$\alpha$} &  ${\rm 1.2\pm0.1}$  & ${\rm3.9\pm0.8}$  &  $33_{-6}^{+8}$  &       $6.5\pm0.1$       &    $318^{+94}_{-90}$               &            --             &  2775 (3076)   \\
                {\sf axill}         &  ${\rm 2.1_{-0.2}^{+0.1}}$  & ${\rm 6.4\pm0.8}$ & $99_{-35}^{+91}$ &       --       &    --     & ${\rm 2.8_{-0.9}^{+1.1}}$ &   2728 (3077)   \\ \hline
                
        \end{tabular}
        \label{tab:models_recap}
        \tablefoot{Column 1: Model name; Col. 2: Photon index; Col. 3: Column density in excess of the Galactic value (units of ${\rm 10^{22}\ cm^{-2}}$); Col. 4: High-energy cutoff rest-frame energy (keV); Cols. 5-6: Rest-frame energy (keV) and equivalent width (eV) of Fe K$\alpha$ emission line; Col. 7: Reflection fraction; Col. 8: $Cstat$ (degrees of freedom ${\rm \nu}$). The line width of the Fe K$\alpha$ is set to $\sigma=0.1$ keV rest frame.  All errors are computed at the 90\% confidence level for one parameter of interest. \textit{Model list}: Model {\sf pl} = {\tt phabs*zphabs*zpo}; Model {\sf acutpl} = {\tt phabs*zphabs*zcutoffpl}; Model {\sf acutpl+K$\alpha$} = {\tt phabs*zphabs*(zcutoffpl+zgauss)}; Model {\sf arefl+K$\alpha$} = {\tt phabs*zphabs*(pexrav+zgauss)}; Model {\sf axill} = {\tt phabs*zphabs*xillver}. All models include Galactic absorption ($N{\rm_H=4.2\times10^{20}\ cm^{-2}}$). }
\end{table*}
\renewcommand{\arraystretch}{1.0}

\subsection{Search for UFO imprints}
 \label{sec:lines}
With its persistent double-velocity component X-ray wind \citepalias{chartas2009}, \apm is a one-of-a-kind object for studying UFOs in high-$z$ AGN. However, 2019 data (see Fig. \ref{fig:residui}) appear not to show the prominent and broad features previously seen in this source. We thus searched our spectra only for narrow ($\sigma=0.1$ keV, rest-frame) emission and absorption features. We applied the blind method of \citet{miniuttifabian2006}, as implemented in \citet{cappi2009}, over the energy range spanned by past UFO events in \apm ($E\simeq7-14 $ keV) using the {\sf axill} best fit as baseline model, also including the \nustar spectra to correctly model the broadband continuum. We then tested our data including {\tt zgauss} components where the blind search and the residuals to model {\sf axill} (Fig. \ref{fig:residui}, panel (e)) showed hints of absorption lines. 

XMM 301 presents no signs of emission or absorption features additional to those already included in {\sf xillver}, while XMM 101 shows hints of a narrow absorption line at $E_{\rm rest}=11.7\pm0.2$ keV ($\Delta Cstat$/$\Delta\nu=9/2$, rest-frame width set to $0.1$ keV). We ran Monte Carlo simulations \citep{protassov2002} to measure the real significance of the 11 keV line, by simulating $10^4$ broadband spectra from our {\tt axill} best fit model (see Table \ref{tab:models_recap}) using the {\tt fakeit} function of {\tt Xspec}. By searching the simulated spectra for the detection of spurious emission and absorption lines, we built the posterior probability distribution of finding a real detection. Applying such a distribution to the line at 11.7 keV in XMM 101, we find that its significance is lower than the 90\% confidence level. Thus, we find that no UFO features are present in the latest X-ray data of APM 08279. 

\section{X-ray reflection in previous observations}
\label{sec:old_data_both}
\label{sec:old_data}
\begin{figure}[!h]
        \centering
        \includegraphics[width=1.0\linewidth
        ]{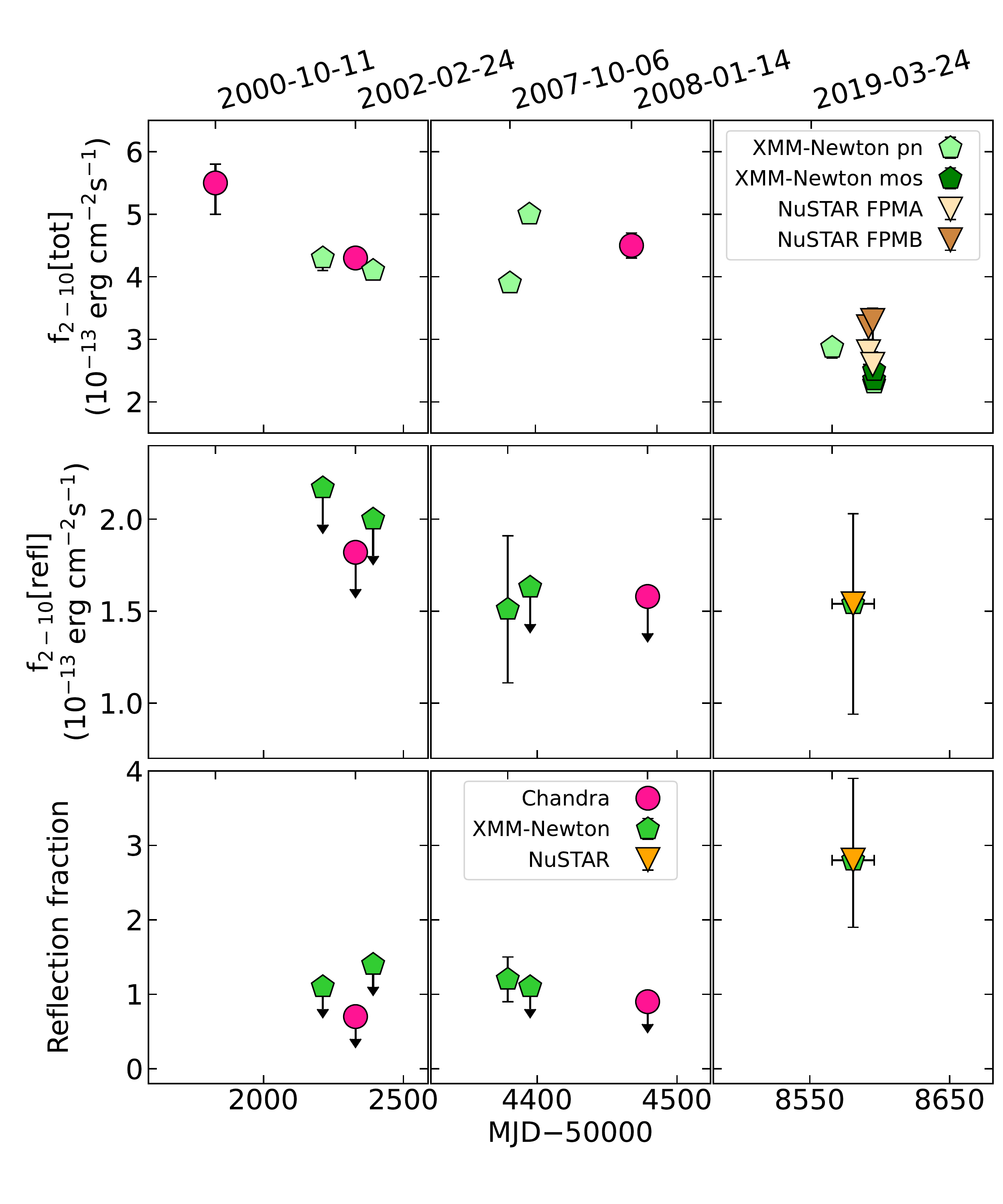}
        \caption{Variation of 2--10 keV observed-band total source flux (top panel), 2--10 keV observed-band flux of the reflection component alone (middle panel) and reflection fraction $R$ (bottom panel),  of \textit{Chandra} (magenta circles), \xmm (light, dark and medium green pentagons for EPIC pn, EPIC MOS, and EPIC pn$+$MOS, respectively) and \nustar (ivory, brown, orange triangles for FPMA, FPMB, and FPMA$+$FPMB, respectively) data from year 2000 to 2019. Top axis shows the dates, bottom axis the modified Julian dates. {\it Top panel}: Values up to 2009 are taken from \citetalias{chartas2009}, the first two points are evaluated from re-analyzed data. Errors are given at 68\% confidence, as in \citetalias{chartas2009}. When they are  not visible, errorbars are smaller than the point size. {\it Middle panel}: Upper limits are computed using the 90\% upper limit of the reflection fraction. 
        {\it Bottom panel}: Upper limits are computed at 90\% confidence level. The first \chandra observation (i.e., the shortest one -- see main text) is left out due to the low count statistics. } 
        \label{fig:flux_r_norm}
\end{figure}

During 2019 exposures, \apm turned out to be fainter than what was previously expected (see Fig. \ref{fig:flux_r_norm}, upper panel, and Tables \ref{tab:red} and \ref{tab:past_obs}), and clearly showed X-ray reflection signatures for the first time. We thus collected all previous \xmm and \chandra X-ray observations with the aim of answering the following questions: namely, whether X-ray reflection was already in place before 2019 and, thus, how that would relate to what was observed in 2019.

By referring to literature studies \citepalias{hasinger2002,chartas2009}, we left out of the data sample the first \chandra exposure, the shortest one, because of the low count statistics. Table \ref{tab:past_obs} summarizes the relevant information about archival observations, including the acronyms that we use to indicate them. We refer to \citetalias{hasinger2002} for the analysis of XMM1 and to \citetalias{chartas2009} for the remaining five (CXO2, XMM2, XMM3, XMM4, CXO3).
We reduced the selected \chandra (CXO2, CXO3) and \xmm (XMM1, XMM2, XMM3, XMM4) archival exposures to uniformly apply the latest calibration files, filtering high-background intervals in \xmm data using the same GTI thresholds of \citetalias{chartas2009}. The count statistics of these spectra is much higher than that of 2019 data, thanks to the higher flux of \apm and to the longer exposures. Similarly to the approach of \citetalias{chartas2009}, spectra with more than $10^4$ cts were grouped to 100 cts/bin (EPIC-pn data of XMM2, XMM3, XMM4), while those with few $10^3$ cts to 20 cts/bin (XMM1 EPIC-pn, all EPIC-MOS and \chandra spectra). We thus applied $\chi^2$ statistics in the spectral analysis of past observations, jointly fitting EPIC-pn and -MOS spectra collected in the same epoch.

\renewcommand{\arraystretch}{1.15}
\begin{table*}[h]
        \begin{center}
                \caption{Log of APM 08279+5255 past observations and reflection parameters}
                \label{tab:past_obs}
                \begin{tabular}{cccccccc}
                        \hline\hline
                               Name         &   ObsID    & Date        &    MJD     & Total Exp. &      $f_{2-10}$      & Ref. &          $R$                     \\ \hline
                                   CXO1            &    1643    & 2000 Oct 11 &   51828    &     9.1     & $5.5_{-0.5}^{+0.3}$  &  a   &          --                     \\
                                   XMM1            & 0092800101 & 2001 Oct 30 &   52212    &    16.5     & $4.3_{-0.2}^{+0.1}$  &  a   & $ <1.1 $    \\
                                   CXO2            &    2979    & 2002 Feb 24 &   52329    &   88.8             & $ 4.3_{-0.1}^{+0.1}$ &  b   &  $<0.7$   \\
                         XMM2              & 0092800201 & 2002 Apr 28 &   52392    &    102.3    & $ 4.1_{-0.1}^{+0.1}$ & a, b & $<1.4$ \\
                                   XMM3            & 0502220201 &  2007 Oct 06 &   54379    &    89.6    & $ 3.9_{-0.1}^{+0.1}$ &  b   &  $1.2_{-0.3}^{+0.3  }$   \\
                                   XMM4            & 0502220301 & 2007 Oct 22 &   54395    &   90.5      & $ 5.0_{-0.1}^{+0.1}$ &  b   & $<1.1$  \\
                                   CXO3            &    7684    & 2008 Jan 14 &   54479    &   88.1             & $ 4.5_{-0.2}^{+0.2}$ &  b   & $<0.9$   \\ \hline
                \end{tabular}
        \end{center}
        \tablefoot{Absorbed fluxes are estimated in the 2--10 keV observed-band from phenomenological models, and are given in units of $10^{-13}$\,erg\,cm$^{-2}$s$^{-1}$, with errors given at 68\% confidence level. The absorbed flux of CXO1 and XMM1 is evaluated from reanalyzed data (see Sect. \ref{sec:old_data}), while the other values are taken from the literature (references: a. \citetalias{hasinger2002}, b. \citetalias{chartas2009}). The total exposure is given in units of ks. The reflection fraction is obtained using {\sf lit\_xill} model), i.e. a {\tt zphabs*zedge*xillver} model for XMM1 and a {\tt zphabs*(xillver+zgauss+zgauss)} model for the other data. Values for the absorber, the edge and the absorption lines are set to the best fit of \citetalias{hasinger2002} and \citetalias{chartas2009}. Errors and upper limits on the reflection fraction are given at the 90\% confidence level.}
\end{table*}    
\renewcommand{\arraystretch}{1.0}
To probe the X-ray reflection, we first drew on the  best-fit models from the literature. \citetalias{hasinger2002} fit XMM1 data with an absorption edge, whereas \citetalias{chartas2009} fit the spectra  firstly using phenomenological models, where the UFOs are modeled only through their main absorption imprint, and secondly using \xstar analytical models ({\tt warmabs}). Testing X-ray reflection while already accounting for UFOs through {\tt warmabs} is a non-trivial exercise, so we selected the phenomenological representation of past data (Model 6, \citetalias{chartas2009}). 
We then tested the X-ray reflection by replacing the single power-law emission of literature phenomenological best-fit models with a {\tt xillver} component ({\sf lit\_xill} model). We set the absorption-edge parameters in XMM1 to those reported by \citetalias{hasinger2002}. Both energy and width of the UFO absorption lines are set to those presented in \citetalias{chartas2009} and we leave their normalization free to vary. 
The lack of high-energy coverage in past observations prevents us from constraining the high-energy cutoff. We thus set it to the value measured in 2019 data ($E_{\rm cut}=100$ keV), safely far from the high-energy end of \chandra and \xmm spectra. We also set the reflection inclination angle to $i=30$\textdegree, as done in Sect. \ref{sec:continuum}, and assume a low-ionization reflecting medium as found for 2019 data ($\rm\log(\xi/erg~s^{-1}~cm )=1.7$). 

We investigated the evolution of the source emission through the epochs using three probes: \textit{i)} the 2--10 keV observed-band total source flux, \textit{ii)} the 2--10 keV observed-band flux of the reflection component only, and \textit{iii)} the reflection fraction $R$. To measure probe \textit{ii)}, we built a pure-reflection model using {\tt xillver} with $R=-1$, setting its normalization to the best-fit value of the respective {\sf lit\_xill} model scaled by the measured reflection fraction. When $R$ is only constrained as an upper limit, we derived an upper limit to the 2--10 keV observed-band reflection flux. When $R$ is constrained (as is for XMM3 and 2019 data), we measured the reflection flux by scaling the normalization of the {\sf lit\_xill} model by the best fit value of $R$. We then assigned as uncertainty the spread in flux obtained by scaling the normalization by the respective upper and lower 90\% confidence level values of $R$. Figure \ref{fig:flux_r_norm} shows the evolution of the three probes across the considered observing epochs. The source X-ray flux (upper panel) presents a factor of $1.5$ decrease between the period 2000--2008 and 2019 while, despite the many upper limits, the reflection flux (middle panel) is fully consistent with showing no trend across the years. 
The 2--10 keV observed-energy band corresponds to $\simeq9.8-49$ keV in the $z=3.91$ rest-frame, which is where X-ray reflection induces the so-called ``reflection hump'' and where flux suppression by absorption is strong only in Compton-thick AGN \citep[e.g.,][]{maiolino1998,bassanidadina1999,matt2000}. Regarding the obscuration hypothesis, we find no evolution in the column density of the low-ionization absorber; the one measured in 2019 data (see Sect. \ref{sec:continuum} and Table \ref{tab:models_recap}) is consistent with results from the literature \citepalias{hasinger2002,chartas2009}. As a consequence, the reduction in flux must be ascribed to a decrease of the source primary activity. 
The trend (or lack of it) in the first two panels of Fig. \ref{fig:flux_r_norm} is indeed well matched by the evolution of $R$ (lower panel) which, despite the large uncertainties, shows a discontinuity between 2019 and epochs prior to 2008. Coupling the three probes together suggests that the reflection component seen in 2019 data was likely already in place before 2008, but less evident due to a stronger primary continuum. 

In this scenario, the reflection component observed in 2019 can possibly be interpreted as the echo of \apm previous activity, due to \textit{(i)} the time delay between the X-ray source and the reflector and \textit{(ii)} the reduced direct emission observed in the last epoch. This is similar to the case of PG 1247+267: \citet{lanzuisi2016} explain its very high reflection fraction in terms of X-ray source variability, namely, the primary emission has dropped but the reflection still has not due to the additional light-travel path. 
Using the same argument, we can place a lower limit on the distance between reflector and X-ray corona, assuming that the variability during the 2008--2019 observational gap is only ascribed to a uniform decrease in the activity of the X-ray corona. We consider the time elapsed between CXO3 and XMM 101 ($\Delta t=832.4$~d in the quasar rest frame) as that corresponding to the light travel path between X-ray source and reflector ($r_{\rm refl}=c\Delta t$). Under this assumption, we derive the lower limit to the reflector location as $r_{\rm refl}\gtrsim0.7$~pc (in accordance to the lower limit obtained from the Fe K$\alpha$ line width: $r_{\rm ~Fe~K\alpha}\gtrsim0.04$ pc), which definitely excludes a disc origin in favor of a distant reflector. By this lower limit, the reflection likely happens in the molecular torus \citep[e.g.,][]{burtscher2013,netzer2015} or, based on the estimate by \citet{saturni2016} for this quasar, at the boundary of the broad-line region. 

\section{Properties of the X-ray corona}
\label{sec:discussion}
We presented a detailed analysis of the first \nustar observations of APM 08279, a gravitationally lensed, broad-absorption line quasar at $z=3.91$, taken jointly to the latest \xmm exposures in 2019. 
By means of primary-emission decrease (see Sect. \ref{sec:old_data}) and high-energy sampling, we are able for the first time to see and constrain a strong reflection component ($R=2.8_{-0.9}^{+1.1}$) and the high-energy cutoff ($E_{\rm cut}=99_{-35}^{+91}$ keV) in this source. Despite the large uncertainties, the high-energy cutoff of \apm is fully consistent with the only other estimates at $z>1$ (\citealt{dadina2016}; \citetalias{lanzuisi2019}; see also the tentative measure of \citealt{lanzuisi2016}). We thus break the previous redshift record of B1422 (\citealt{dadina2016}; \citetalias{lanzuisi2019}) and find additional evidence for complex emission mechanisms, very much alike those of local Seyfert galaxies, in high-$z$ AGN (up to $z\approx4$). 
It is of interest to notice that we do not find evidence for significant X-ray winds in the 2019 observations of APM 08279, the archetype of high-$z$ UFOs. While these winds are known to be variable and episodic events \citep[e.g.,][]{dadina2005,cappi2006,giustini2011,gofford2014,igo_2020,parker2021}, the present data do not allow us to investigate further on their disappearance. 
Only a new, dedicated monitoring will be key in probing  \textit{i)}  whether this new flux state is enduring, \textit{ii)} whether UFOs are no longer a distinctive feature of \apm, and \textit{iii)} whether the former might be the cause of the latter. 

\begin{figure*}[h]
        \centering
        \includegraphics[width=0.49\linewidth
        ]{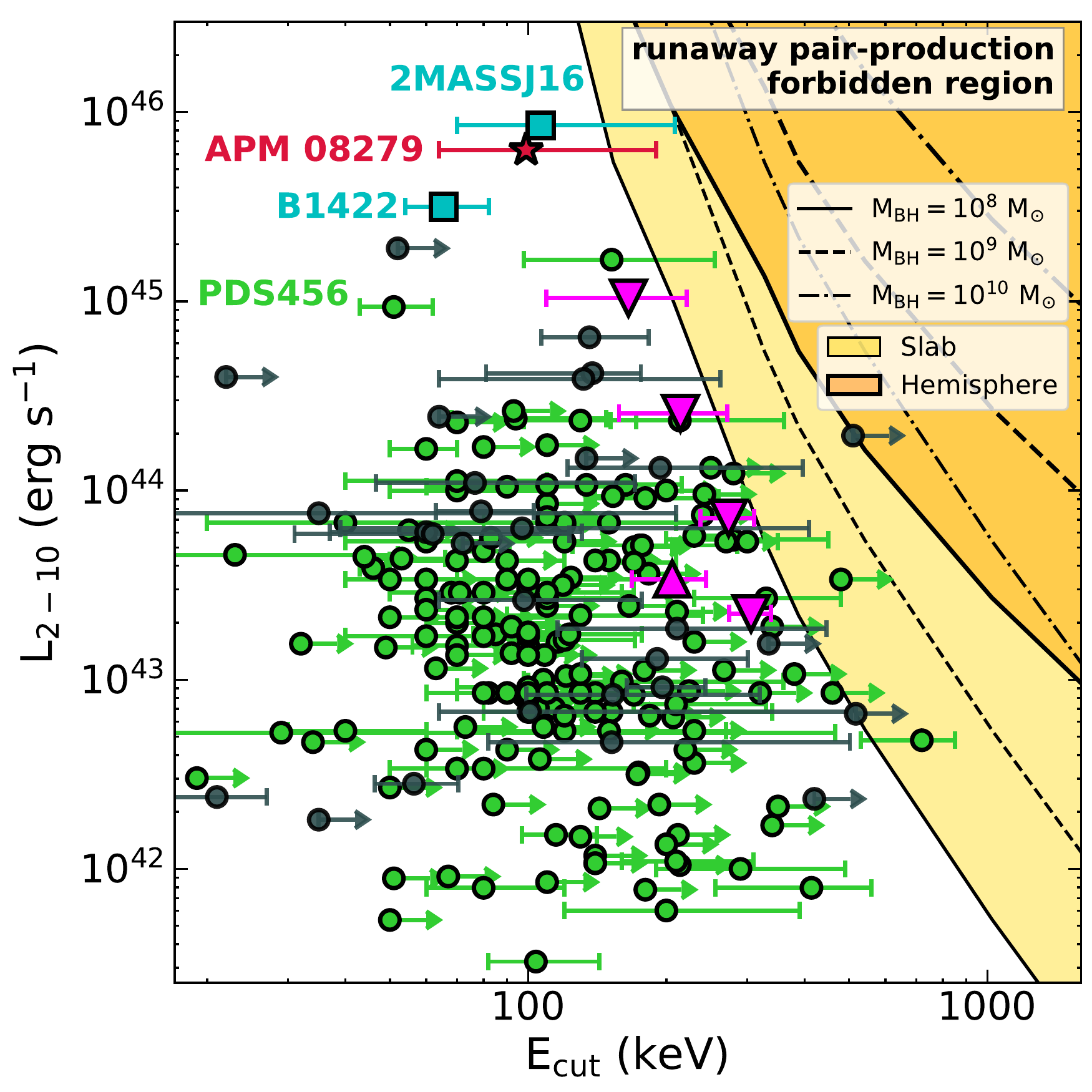}
        \hfill
        \includegraphics[width=0.49\linewidth
        ]{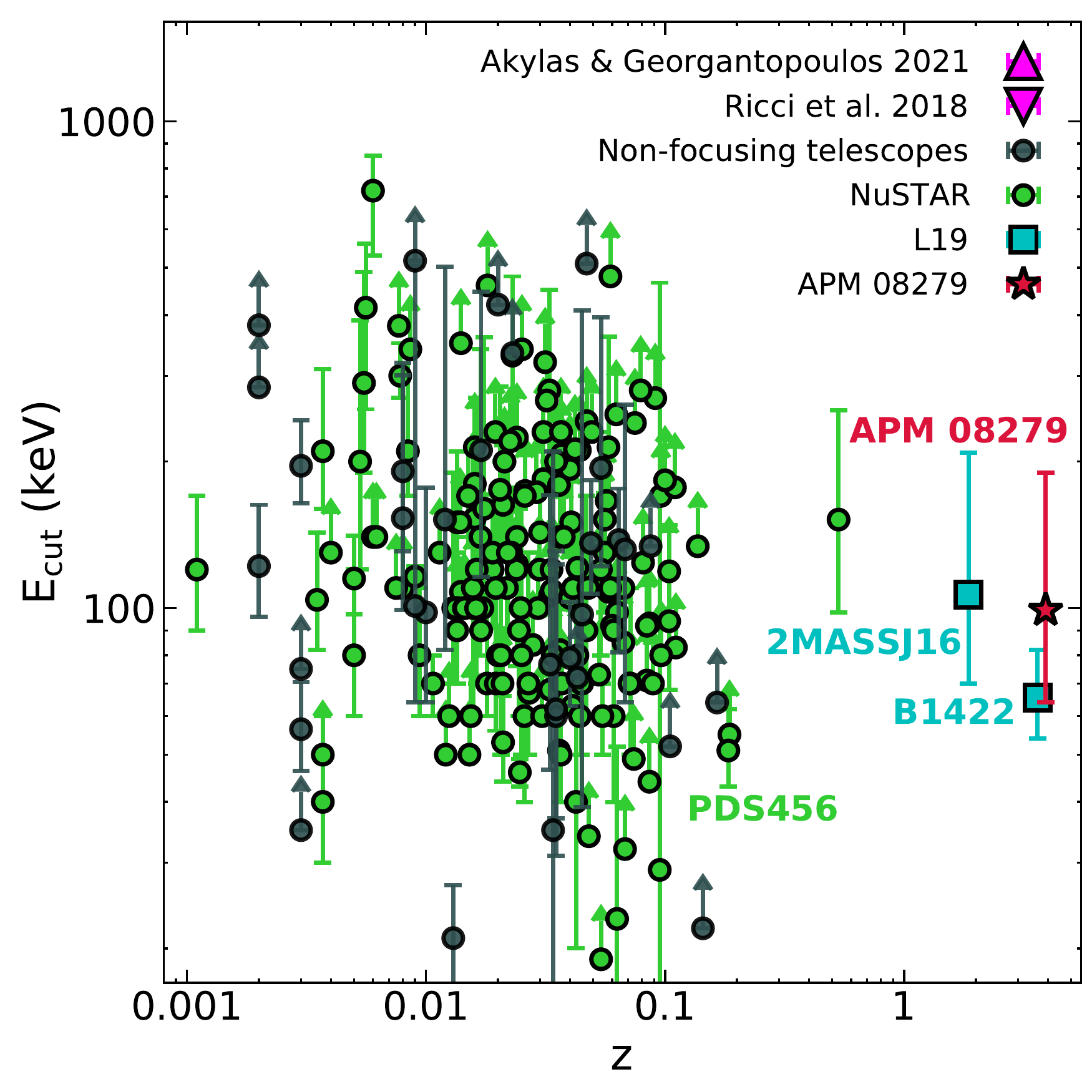}
        \caption{Compactness--temperature diagram translated into directly observable quantities and high-energy cutoff distribution as a function of redshift. 
        \textit{Left:} X-ray luminosity vs. high-energy cutoff, updated from \citetalias{lanzuisi2019} (see main text and Tables \ref{tab:ecutplot1}--\ref{tab:ecutplot3} for details). \nustar measurements are in green, non-focusing telescopes' in grey, high-$z$ AGN from \citetalias{lanzuisi2019} in cyan, and our measurement in red. Magenta downward triangles mark the averaged values for BASS AGN from \citet{ricci2018}, while the magenta upward triangle marks the median point of Seyfert 1 galaxies from \citet{akylas2021}. Yellow (orange) areas delimited by a thin (thick) line show the runaway pair-production region for a $10^8~M_{\odot}$ SMBH in the case of slab (hemisphere) geometry.
        Dashed (dashed-dotted) lines mark the same thresholds for a $10^9~M_{\odot}$ ($10^{10}~M_{\odot}$) SMBH. \textit{Right: } High-energy cutoff vs. redshift plane for the same samples as in the left panel. }
        \label{fig:cfr_lanz19}
\end{figure*}

The measured reflection fraction ($R=2.8_{-0.9}^{+1.1}$) is, even considering the large uncertainties, much higher than what expected based on previous results in the literature for high-luminosity sources: AGN with $L_{\rm X}>10^{45}$ $\rm erg~s^{-1}$ are usually found to show $R<1$ \citep{delmoro2017,zappacosta2018}. 
We note that even though lower than our estimate, both the reflection fractions measured in \citetalias{lanzuisi2019} are higher than what expected for high-luminosity sources. However, the case of \apm during 2019 exposures quite differs from those of B1422 and 2MASSJ16, and is more alike that of PG 1247+267, as discussed in Sect. \ref{sec:old_data}. 
Moreover, the Fe K$\alpha$ EW of APM 08279, measured with the phenomenological model {\sf acutpl+K$\alpha$} (Table \ref{tab:models_recap}), is larger than the expectation based on the results over samples of high-$z$ and local AGN \citep[e.g.,][]{falocco2013} and on the Iwasawa-Taniguchi effect \citep{bianchi2007}, but is consistent, at 90\% confidence level, with the highest EWs of the CAIXA sample \citep[corresponding to the 98\% percentile of the EW distribution,][]{bianchi2009}. 
In fact, such high values of $R$ and of Fe K$\alpha$ EW in 2019 data, coupled with the non-evolution of the X-ray reflection flux in the 2--10 keV observed band and the lower limit placed on the reflector's distance, likely hint that the reflection component in this last observation could be the echo of \apm previous activity. Moreover, the majority of the reflection fraction values obtained from data prior to 2008 (see Fig. \ref{fig:flux_r_norm} and Table \ref{tab:past_obs}) are in decent agreement with what expected from the literature ($R<1$). This is additional proof for our interpretation of \apm activity variation (see Sect. \ref{sec:old_data}), and for the Seyfert-like mechanisms that give origin to its emission. 

\citetalias{lanzuisi2019} adopted the $\ell-\theta$ plane of \citet{fabian2015} translated into the directly observable quantities $L_{\rm X}$ versus $E_{\rm cut}$, respectively, through Eqs. \eqref{eq:theta} (with $K=2$, as assumed in \citealp{fabian2015}) and \eqref{eq:compactness}. In the same way, theoretical $\ell$($\theta$) critical lines of \citet{stern1995} were converted into $L_{\rm X}$($E_{\rm cut}$) thresholds, assuming a corona size $R_{\rm X}=5\,r_{\rm S}$ (Schwarzschild radius: $r_{\rm S}=2GM_{\rm BH}/c^2$) and SMBH masses $M_{\rm BH}=10^8\,M_{\odot}$ and $10^9\,M_{\odot}$, namely, representative of the sample of \citet{fabian2015} and of powerful, high-$z$ quasars, respectively.
\citetalias{lanzuisi2019} also updated the compilation by \citet{fabian2015} with more recent $E_{\rm cut}$ measurements, obtained both with and without \nustar data, and both on individual targets or over samples of sources \citep[e.g.,][]{malizia2014,tortosa2017,buisson2018,molina2019}. \citetalias{lanzuisi2019} included also the median values, binned in compactness (i.e., $L_{\rm X}$), of the large BASS sample of local AGN as measured by \citet{ricci2018}. Figure \ref{fig:cfr_lanz19} (left) shows the $L_{\rm X}$--$E_{\rm cut}$ plane of \citetalias{lanzuisi2019} further updated with our measurement, the pair-production critical lines for $M_{\rm BH}=10^{10}\,M_{\odot}$, and results from other recent works, derived through canonical modeling of the continuum \citep[][see also \citealp{ezhikode2020}, where the authors model X-ray reflection accounting for relativistic effects]{kara2017,tortosa2018,kamraj2018,ursini2020,balokovic2020,middei2020,middei2021,reeves2021}. When multiple estimates for a same source are available, we only kept the latest one. We also added earlier measurements of \citet{dadina2007} that were not revised in later works, and selecting the best constrained value in case of multiple measurements for a single source. 
\citet{akylas2021} recently studied the distribution of coronal temperatures in a big sample of Seyfert 1 selected by {\it Swift} and followed up by \nustar (118 sources, many also comprised in \citealt{ricci2017,ricci2018}). Figure \ref{fig:cfr_lanz19} shows the median high-energy cutoff of Seyfert 1 galaxies measured by \citet{akylas2021}.
Recent works by \citet{saturni2016,saturni2018} have agreed on \apm data being best reproduced by low magnification factors ($\mu_{\rm L}\lesssim9$). To add our target to the $L_{\rm X}$--$E_{\rm cut}$ plane, we estimated the de-absorbed and de-lensed 2--10 keV luminosity of \apm ($L_{\rm X}=6.5\times10^{45}\rm\, erg\,s^{-1}$) assuming a magnification factor of $\mu_{\rm L}=4$ \citep{riechers2009}. 
Like the other high-$z$ AGN, \apm falls in the allowed region for high-mass SMBHs with X-ray luminosity higher than $2\times10^{45}\rm~erg\, s^{-1}$. 
Being located in the proximity of the central SMBH, hot coronae are subject to the laws of general relativity \citep[e.g.,][]{wilkins2021}. \citet{tamborra2018} provide correction factors to be applied to results obtained through canonical reflection models. The authors demonstrate that, without accounting for general relativity, the observed high-energy cutoff ($E^o_{\rm cut}$) underestimates the intrinsic value $E^i_{\rm cut}$ by a factor $g=E^i_{\rm cut}/E^o_{\rm cut}$. \citet{tamborra2018} compute g-factors for a variety of combinations of corona properties and reflection inclination angles. Assuming a $5\,r_{\rm S}$ corona and an inclination angle $i=30$\textdegree~(see Sect. \ref{sec:continuum}), the corresponding g-factor spans between 1.2 and 1.5. Even assuming the maximum value, the 90\% confidence level upper bound of the high-energy cutoff of \apm falls below the critical line for a slab corona and $M_{\rm BH}=10^{10}~M_{\odot}$. Thus, \apm would still lie in the allowed region of the $L_{\rm X}-E_{\rm cut}$ plane for a SMBH mass of $\simeq10^{10}~M_{\odot}$ even when general relativity effects are accounted for. 

\citetalias{lanzuisi2019} found that the median cutoff energy expected for local BASS AGN \citep{ricci2018} in the same accretion regime as their high-$z$ sources was much higher than the measured values of B1422 and 2MASSJ16. 
For what concerns APM 08279, local AGN in the same Eddington regime as our target ($\lambda_{\rm Edd}\simeq0.4$, \citealt{saturni2018}) show a median high-energy cutoff of $E_{\rm cut}\simeq170$~keV, which is well above the best-fit measurement ($E_{\rm cut}\approx99$ keV) but consistent with its 90\% confidence range ($68\ {\rm keV}<E_{\rm cut}<190$ keV). However, the existence of a relation between high-energy cutoff and Eddington ratio was recently debated in the literature: \citet{hinkle2021} and \citet{kamraj2022} find no correlation between accretion parameters and the high-energy cutoff in their new analyses of BASS AGN, as opposed to what seen by \citet{ricci2018}. According to their spectral analysis of \nustar data alone of \textit{Swift}/BAT selected AGN, \citet{kang2022} confirmed the absence of a $E_{\rm cut}$--$\lambda_{\rm Edd}$ relation and, interestingly, they find that some sources fall in the runaway pair-production region of Fig. \ref{fig:cfr_lanz19}.
Only a better sampling of both the high-luminosity end and the high-accretion regime will allow us to better understand the physics regulating hot coronae in powerful AGN. 
Lastly, we addressed the possibility of \apm  having a magnification factor more similar to $\mu_{\rm L}=100$ \citep{egami2000}, which would make our target a more "regular" AGN ($L_{\rm X}(\mu_{\rm L}=100)=2.6\times10^{44}\rm\, erg\,s^{-1}$, $\lambda_{\rm Edd}(\mu_{\rm L}=100)\simeq0.08$). Nevertheless, \apm would still fall in the allowed region of the $L_{\rm X}-E_{\rm cut}$ plane and its high-energy cutoff would not be consistent with the median value of BASS AGN in a similar accretion regime.  

\section{Summary and conclusions}
\label{sec:summary}
We presented our analysis of the first X-ray broadband spectrum of APM 08279, a gravitationally lensed, broad-absorption line quasar at $z=3.91$. We then compared our findings with past observations and we summarize our key results here: 
\begin{itemize}
    \item We measured a long-lasting X-ray reflection component in this source for the first time. We find it consistent with being produced by distant material, in the molecular torus or at the boundary of the broad-line region ($r_{\rm refl}\gtrsim0.7~{\rm pc}$). The large reflection fraction of $R\simeq2.8_{-0.9}^{+1.1}$ is interpreted in terms of a decrease in the primary X-ray emission, as, for instance, the case of PG 1247+267.
    \item We break the previous redshift record for the farthest high-energy cutoff ever observed. Our measurement of $E_{\rm cut}=99_{-35}^{+91}$~keV is fully in agreement with previous results on high-$z$, high-luminosity AGN.
    \item As opposed to the other high-$z$ sources of \citetalias{lanzuisi2019}, the cutoff measured for \apm is consistent (within a 90\% confidence level) with the median values of BASS AGN in the same accretion regime. A better sampling of this luminosity regime will be key to study the behavior of such sources. 
\end{itemize}
Results that have come about over the last two decades have shown that studies of high-$z$ AGN are observationally challenging but highly rewarding, when carried out by means of long-enough exposures of present-day observatories. Next-generation X-ray telescopes will be crucial in substantially expanding the samples used to test the physical processes that regulate hot coronae. In particular, enlarging the sample of high-$z$ AGN, especially at the fainter end, will shed light on whether low- and high-$z$ AGN comply with the same relations. Would they differ, only future studies of high-$z$ AGN will disclose whether it is owed to the different luminosity regimes or whether it is a byproduct of a potential evolution with cosmic time. To this aim, the \textit{eROSITA} All-Sky Survey will be key in discovering new high-$z$ AGN, to then be followed up by present facilities such as \nustar and, hopefully, future hard X-ray instruments.

\begin{acknowledgements}
We acknowledge financial support from ASI under grants ASI-INAF I/037/12/0 and n. 2017-14-H.O, and from the grant PRIN MIUR 2017PH3WAT (‘Black hole winds and the baryon life cycle of galaxies’). 
This work is based on observations obtained with XMM-\textit{Newton}, an ESA science mission with instruments and contributions directly funded by ESA Member States and the USA (NASA). This research has made use of data obtained from the \textit{Chandra} Data Archive, and software provided by the \textit{Chandra} X-ray Center (CXC) in the application package CIAO. 
This research has made use of data and/or software provided by the High Energy Astrophysics Science Archive Research Center (HEASARC), which is a service of the Astrophysics Science Division at NASA/GSFC. This work made use of data from the \textit{NuSTAR} mission, a
project led by the California Institute of Technology, managed by the Jet Propulsion Laboratory, and funded by the NASA. This research has made use of the NuSTAR Data Analysis Software (NuSTARDAS) jointly developed by the ASI Science Data Center (ASDC, Italy) and the California Institute of Technology (USA).

\end{acknowledgements}

\bibliographystyle{aa}
\bibliography{apm08279_acc}
\clearpage
\onecolumn
\begin{appendix}
\section{High-energy cutoff measurements by \nustar and nonfocusing telescopes}
\begin{longtable}{lccccc}
\caption{Literature low-$z$ measurements collected by \citetalias{lanzuisi2019}}
\label{tab:ecutplot1}\\
\hline\hline
Source & $z$ &  $\log L_{2-10}$  &  $E_{\rm cut}$  & Flag & References   \\
\hline
\endfirsthead

\caption*{continued.}\\
\hline\hline
Source & $z$ &  $\log L_{2-10}$  & $E_{\rm cut}$  & Flag & References   \\
\hline
\endhead

\hline
\endfoot

NGC~5506	   & 0.006  &  42.68 & $720_{-190}^{+130}$ & 1    & (a),(b)   \\ 
NGC~7213	   & 0.006  &  42.07 & $<140$		   & 1    & (a),(b)   \\ 
MCG-6-30-15	   & 0.008  &  43.39 & $<110$		   & 1    & (a),(b)   \\ 
MCG~5-23-16	   & 0.009  &  43.2  & $116_{-5}^{+6}$     & 1    & (a),(b)   \\ 
SWIFT~J2127.4+5654 & 0.014  &  43.0  & $108_{-10}^{+11}$   & 1    & (a),(b)   \\ 
NGC~5548	   & 0.018  &  43.3  & $70_{-10}^{+40}$    & 1    & (a),(b)   \\ 
Mrk~335 	   & 0.026  &  42.51 & $<174$		   & 1    & (a),(b)   \\ 
1H0707-495	   & 0.041  &  43.06 & $<63$		   & 1    & (a),(b)   \\ 
Fairall~9	   & 0.047  &  43.98 & $<242$		   & 1    & (a),(b)   \\ 
Cyg~A		   & 0.056  &  44.24 & $<110$		   & 1    & (a),(b)   \\ 
3C~382  	   & 0.058  &  44.37 & $214_{-63}^{+147}$  & 1    & (a),(b)   \\ 
IGR~J0033+6122     & 0.105  &  45.28 & $<52$		   & 2    & (c),(a),(b)  \\ 
3C~111  	   & 0.049  &  44.81 & $136_{-29}^{+47}$   & 2    & (c),(a),(b)  \\ 
IGR~J07597-3842    & 0.04   &  43.89 & $79_{-16}^{+24}$    & 2    & (c),(a),(b)  \\ 
NGC~3783	   & 0.01   &  43.42 & $98_{-34}^{+79}$    & 2    & (c),(a),(b)  \\ 
NGC~4151	   & 0.003  &  42.96 & $196_{-32}^{+47}$   & 2    & (c),(a),(b)  \\ 
IGR~J16558-5203    & 0.054  &  44.12 & $194_{-72}^{+202}$  & 2    & (c),(a),(b)  \\ 
1H2251-179	   & 0.064  &  44.62 & $138_{-57}^{+38}$   & 2    & (c),(a),(b)  \\ 
MCG-02-58-022	   & 0.047  &  44.29 & $<510$		   & 2    & (c),(a),(b)  \\ 
MCG+08-11-011	   & 0.021  &  44.03 & $163_{-32}^{+53}$   & 1    & (d),(b)   \\ 
Mrk~6		   & 0.019  &  43.21 & $120_{-28}^{+51}$   & 1    & (d),(b)   \\ 
IGR~J12415-5750    & 0.024  &  43.24 & $123_{-47}^{+54}$   & 1    & (d),(b)   \\ 
IC4329A 	   & 0.016  &  43.97 & $153_{-16}^{+20}$   & 1    & (d),(b)   \\ 
GRS~1734-292	   & 0.021  &  43.64 & $53_{-9}^{+13}$     & 1    & (d),(b)   \\ 
3C~390.3	   & 0.056  &  44.37 & $130_{-32}^{+42}$   & 1    & (d),(b)   \\ 
NGC~6814	   & 0.005  &  42.18 & $115_{-18}^{+26}$   & 1    & (d),(b)   \\ 
4C~74.24	   & 0.104  &  44.38 & $94_{-26}^{+54}$    & 1    & (d),(b)   \\ 
S5~2116+81	   & 0.086  &  44.42 & $<93$		   & 1    & (d),(b)   \\ 
PG~1114+445	   & 0.144  &  44.6  & $<22$		   & 2    & (e),(a),(b)\\
NGC~4051	   & 0.002  &  40.68 & $<381$		   & 2    & (e),(a),(b)\\
PG~1202+281	   & 0.165  &  44.39 & $<64$		   & 2    & (e),(a),(b)\\
NGC~4138	   & 0.003  &  41.21 & $<75$		   & 2    & (e),(a),(b)\\
Mrk~766 	   & 0.013  &  42.38 & $21_{-7}^{+6}$	   & 2    & (e),(a),(b)\\
NGC~4258	   & 0.002  &  40.82 & $<284$		   & 2    & (e),(a),(b)\\
Mrk~50  	   & 0.023  &  43.19 & $<334$		   & 2    & (e),(a),(b)\\
NGC~4593	   & 0.009  &  42.82 & $<517$		   & 2    & (e),(a),(b)\\
Mrk~1383	   & 0.087  &  44.17 & $<134$		   & 2    & (e),(a),(b)\\
NGC~3998	   & 0.0035 &  41.51 & $104_{-22}^{+39}$   & 1    & (f),(b)    \\
NGC~4579	   & 0.0056 &  41.9  & $414_{-158}^{+146}$ & 1    & (f),(b)    \\
ESO~103-035	   & 0.013  &  42.96 & $100_{-30}^{+90}$   & 1    & (g),(b)   \\
IGR~2124	   & 0.02   &  43.68 & $80_{-9}^{+11}$     & 1    & (g),(b)   \\
B2202-209	   & 0.532  &  45.22 & $152_{-54}^{+103}$  & 1    & (h),(b)  \\ 
GRS~1734-292	   & 0.021  &  43.72 & $53_{-8}^{+11}$     & 1    & (i),(b)  \\ 
   \end{longtable}
   \tablefoot{Col. (1): Source name; Col. (2): Redshift; Col. (3): Logarithm of the 2--10 keV luminosity ($\rm erg~s^{-1}$); Col. (4): High-energy cutoff (keV); Col. (5): telescope flag, 1={\it NuSTAR}, 2=nonfocusing; Col. (6): References. }
 \tablebib{(a) \citet{fabian2015}, (b) \citetalias{lanzuisi2019}, (c) \citet{malizia2014}, (d) \citet{molina2019}, (e) \citet{vasudevan2013}, (f) \citet{younes2019}, (g) \citet{buisson2018}, (h) \citet{kammoun2017}, (i) \citet{tortosa2017}. }

\newpage
\begin{longtable}{lccccc}
\caption{Literature low-$z$ measurements collected in this work}
\label{tab:ecutplot2}\\
\hline\hline
Source & $z$ &  $\log L_{2-10}$  &  $E_{\rm cut}$  & Flag & References   \\
\hline
\endfirsthead

\caption*{continued.}\\
\hline\hline
Source & $z$ &  $\log L_{2-10}$  &  $E_{\rm cut}$  & Flag & References   \\
\hline
\endhead

\hline
\endfoot
NGC~985 	   & 0.043  &  43.72 & $<72$		   & 2    & (j)    \\
ESO198-G24	   & 0.045  &  43.8  & $97_{-58}^{+312}$   & 2    & (j)    \\
NCG~1068	   & 0.003  &  42.26 & $<35$		   & 2    & (j)    \\
3C~120  	   & 0.033  &  44.04 & $77_{-30}^{+94}$    & 2    & (j)    \\
H0557		   & 0.034  &  43.88 & $35_{-20}^{+175}$   & 2    & (j)    \\
MCG-1-24-12	   & 0.02   &  42.37 & $<420$		   & 2    & (j)    \\
MCG-5-23-16	   & 0.008  &  43.11 & $191_{-60}^{+110}$  & 2    & (j)    \\
NGC~3516	   & 0.009  &  42.83 & $101_{-37}^{+404}$  & 2    & (j)    \\
NGC~4151*	    & 0.003  &  42.45 & $56_{-10}^{+14}$    & 2    & (j)    \\
NGC~4507	   & 0.012  &  42.67 & $152_{-70}^{+350}$  & 2    & (j)    \\
NGC~4945	   & 0.002  &  40.51 & $122_{-26}^{+41}$   & 2    & (j)    \\
Mrk~509*	    & 0.035  &  43.78 & $60_{-23}^{+71}$    & 2    & (j)    \\
MR~2251 	   & 0.068  &  44.59 & $132_{-68}^{+130}$  & 2    & (j)    \\
NGC~7469	   & 0.017  &  43.27 & $211_{-95}^{+235}$  & 2    & (j)    \\
Ark~564 	   & 0.02468&  43.59 & $46_{-3}^{+3}$	   & 1    & (k)        \\ 
MGC~+8-11-11	   & 0.0204 &  43.71 & $175_{-50}^{+110}$  & 1    & (l)     \\ 
Ark~120 	   & 0.033  &  43.96 & $180_{-40}^{+80}$   & 1    & (l)     \\ 
PG~1211+143	   & 0.0809 &  43.54 & $<124$		   & 1    & (l)     \\ 
1E~0754.6+3928     & 0.096  &  43.7  & $<170$		   & 1    & (m)      \\ 
HE~1143-1810	   & 0.0328 &  43.74 & $280_{-80}^{+170}$  & 1    & (n)      \\
MCG-01-24-12	   & 0.0196 &  43.18 & $70_{-14}^{+21}$    & 1    & (o)      \\
PDS~456 	   & 0.184  &  44.97 & $51_{-8}^{+11}$     & 1    & (p)      \\ 
1RXSJ034704.9	   & 0.095  &  42.72 & $29_{-18}^{+437}$   & 1    & (q)      \\
1RXS~J174538.1     & 0.111  &  43.75 & $<83$		   & 1    & (q)      \\
1RXS~J213445.2     & 0.067  &  43.24 & $<85$		   & 1    & (q)      \\
2MASS~J19334715    & 0.057  &  43.39 & $<166$		   & 1    & (q)      \\
2MASX~J04372814    & 0.053  &  42.85 & $<114$		   & 1    & (q)      \\
2MASX~J12313717    & 0.028  &  42.34 & $<84$		   & 1    & (q)      \\
2MASX~J15144217    & 0.068  &  43.19 & $<32$		   & 1    & (q)      \\
2MASX~J15295830    & 0.104  &  43.5  & $<119$		   & 1    & (q)      \\
2MASX~J19301380    & 0.063  &  43.66 & $23_{-9}^{+29}$     & 1    & (q)      \\
2MASX~J19380437    & 0.04   &  42.85 & $<105$		   & 1    & (q)      \\
2MASX~J20005575    & 0.037  &  42.8  & $<207$		   & 1    & (q)      \\
3C~227  	   & 0.086  &  43.65 & $<44$		   & 1    & (q)      \\
4C~+18.51	   & 0.186  &  43.79 & $<55$		   & 1    & (q)      \\
ESO~438-G009	   & 0.024  &  42.03 & $<140$		   & 1    & (q)      \\
Fairall~1146	   & 0.031  &  42.81 & $<184$		   & 1    & (q)      \\
Fairall~1203	   & 0.058  &  42.75 & $<108$		   & 1    & (q)      \\
$[$HB89$]$~0241+622    & 0.044  &  43.36 & $<211$		   & 1    & (q)      \\
IGR~J14471-6414    & 0.053  &  42.75 & $<73$		   & 1    & (q)      \\
IGR~J14552-5133    & 0.016  &  41.89 & $<180$		   & 1    & (q)      \\
IRAS~04392-2713    & 0.084  &  43.46 & $<71$		   & 1    & (q)      \\
LCRSB~232242.2     & 0.036  &  41.95 & $<51$		   & 1    & (q)      \\
Mrk~9		   & 0.04   &  42.34 & $<193$		   & 1    & (q)      \\
Mrk~376 	   & 0.056  &  42.83 & $<152$		   & 1    & (q)      \\
Mrk~595 	   & 0.027  &  41.96 & $<67$		   & 1    & (q)      \\
Mrk~732 	   & 0.029  &  42.5  & $<173$		   & 1    & (q)      \\
Mrk~739 	   & 0.03   &  42.32 & $<143$		   & 1    & (q)      \\
Mrk~813 	   & 0.11   &  43.71 & $<177$		   & 1    & (q)      \\
Mrk~817 	   & 0.031  &  42.56 & $<230$		   & 1    & (q)      \\
Mrk~841 	   & 0.036  &  43.05 & $<179$		   & 1    & (q)      \\
Mrk~1018	   & 0.042  &  42.18 & $<212$		   & 1    & (q)      \\
Mrk~1044	   & 0.016  &  42.02 & $<214$		   & 1    & (q)      \\
Mrk~1310	   & 0.019  &  42.17 & $<130$		   & 1    & (q)      \\
Mrk~1393	   & 0.054  &  42.48 & $<19$		   & 1    & (q)      \\
NGC~0985	   & 0.043  &  43.02 & $<121$		   & 1    & (q)      \\
PG~0804+761	   & 0.1    &  43.56 & $<183$		   & 1    & (q)      \\
PKS~0558-504	   & 0.137  &  44.03 & $<134$		   & 1    & (q)      \\
RBS~0295	   & 0.074  &  43.17 & $<49$		   & 1    & (q)      \\
RBS~0770	   & 0.032  &  43.05 & $<267$		   & 1    & (q)      \\
RBS~1037	   & 0.084  &  43.14 & $<92$		   & 1    & (q)      \\
RBS~1125	   & 0.063  &  42.91 & $<98$		   & 1    & (q)      \\
SBS~1136+594	   & 0.06   &  43.28 & $<92$		   & 1    & (q)      \\
SDSS~J104326.47    & 0.048  &  42.67 & $<34$		   & 1    & (q)      \\
UM~614  	   & 0.033  &  42.58 & $<106$		   & 1    & (q)      \\
WKK~1263	   & 0.024  &  42.94 & $<224$		   & 1    & (q)      \\
NGC~262 	   & 0.015  &  43.62 & $170_{-30}^{+40}$   & 1    & (r)   \\
ESO~195-IG021	   & 0.0494 &  43.76 & $<230$		   & 1    & (r)   \\
NGC~454~E	   & 0.0121 &  42.43 & $<50$		   & 1    & (r)   \\
NGC~513 	   & 0.0195 &  42.73 & $<230$		   & 1    & (r)   \\
NGC~612 	   & 0.0298 &  43.82 & $<120$		   & 1    & (r)   \\
2MASX~J0140	   & 0.0716 &  44.0  & $70_{-20}^{+40}$    & 1    & (r)   \\
MCG-01-05-047	   & 0.0172 &  42.9  & $<100$		   & 1    & (r)   \\
NGC~788 	   & 0.0136 &  43.18 & $<100$		   & 1    & (r)   \\
ESO~416-G002	   & 0.0591 &  43.53 & $<480$		   & 1    & (r)   \\
NGC~1052	   & 0.005  &  41.9  & $80_{-20}^{+40}$    & 1    & (r)   \\
2MFGC~2280	   & 0.0152 &  43.33 & $<50$		   & 1    & (r)   \\
NGC~1229	   & 0.0363 &  42.93 & $<82$		   & 1    & (r)   \\
NGC~1365	   & 0.0055 &  42.0  & $290_{-100}^{+200}$ & 1    & (r)   \\
2MASX~J0356	   & 0.0748 &  43.87 & $<240$		   & 1    & (r)   \\
3C~105  	   & 0.089  &  44.36 & $<70$		   & 1    & (r)   \\
2MASX~J0423	   & 0.045  &  44.05 & $70_{-30}^{+40}$    & 1    & (r)   \\
MCG+03-13-001	   & 0.0154 &  42.63 & $<60$		   & 1    & (r)   \\
CGCG~420-015	   & 0.0294 &  43.48 & $<100$		   & 1    & (r)   \\
ESO~033-G002	   & 0.0181 &  42.93 & $<460$		   & 1    & (r)   \\
LEDA~178130	   & 0.035  &  44.0  & $<200$		   & 1    & (r)   \\
2MASX~J0508	   & 0.0175 &  42.99 & $160_{-60}^{+200}$  & 1    & (r)   \\
NGC~2110	   & 0.0078 &  43.73 & $300_{-30}^{+50}$   & 1    & (r)   \\
ESO~005-G004	   & 0.0062 &  42.03 & $<140$		   & 1    & (r)   \\
Mrk~3		   & 0.0135 &  43.83 & $150_{-30}^{+60}$   & 1    & (r)   \\
ESO~121-IG028	   & 0.0405 &  43.63 & $<150$		   & 1    & (r)   \\
LEDA~549777	   & 0.061  &  43.63 & $<90$		   & 1    & (r)   \\
LEDA~511628	   & 0.0469 &  43.53 & $90_{-30}^{+80}$    & 1    & (r)   \\
MCG+06-16-028	   & 0.0157 &  42.93 & $<110$		   & 1    & (r)   \\
IRAS~07378-3136    & 0.0258 &  43.23 & $60_{-20}^{+40}$    & 1    & (r)   \\
UGC~3995~A	   & 0.0158 &  42.93 & $100_{-40}^{+110}$  & 1    & (r)   \\
Mrk~1210	   & 0.0135 &  42.93 & $90_{-20}^{+40}$    & 1    & (r)   \\
MCG-01-22-006	   & 0.0218 &  43.43 & $110_{-30}^{+60}$   & 1    & (r)   \\
CGCG~150-014	   & 0.0647 &  43.93 & $<110$		   & 1    & (r)   \\
MCG+11-11-032	   & 0.0363 &  43.63 & $<140$		   & 1    & (r)   \\
2MASX~J0903	   & 0.091  &  43.73 & $<270$		   & 1    & (r)   \\
2MASX~J0911	   & 0.0268 &  43.33 & $70_{-20}^{+60}$    & 1    & (r)   \\
IC~2461 	   & 0.0075 &  41.93 & $<110$		   & 1    & (r)   \\
MCG-01-24-012	   & 0.0196 &  43.46 & $110_{-30}^{+50}$   & 1    & (r)   \\
2MASX~J0923	   & 0.0424 &  43.83 & $40_{-20}^{+90}$    & 1    & (r)   \\
NGC~2992	   & 0.0077 &  43.03 & $<380$		   & 1    & (r)   \\
NGC~3079	   & 0.0037 &  42.73 & $40_{-10}^{+20}$    & 1    & (r)   \\
ESO~263-G013	   & 0.0335 &  43.73 & $<120$		   & 1    & (r)   \\
NGC~3281	   & 0.0107 &  42.53 & $70_{-10}^{+10}$    & 1    & (r)   \\
MCG+12-10-067	   & 0.0336 &  43.13 & $<109$		   & 1    & (r)   \\
MCG+06-24-008	   & 0.0259 &  42.92 & $<170$		   & 1    & (r)   \\
UGC~5881	   & 0.0206 &  42.53 & $80_{-30}^{+120}$   & 1    & (r)   \\
NGC~3393	   & 0.0125 &  43.53 & $<60$		   & 1    & (r)   \\
2MASX~J1136	   & 0.014  &  42.33 & $<350$		   & 1    & (r)   \\
NGC~3822	   & 0.0209 &  42.53 & $<70$		   & 1    & (r)   \\
B2~1204+34	   & 0.0791 &  44.09 & $<280$		   & 1    & (r)   \\
IRAS~12074-4619    & 0.0315 &  42.93 & $<320$		   & 1    & (r)   \\
WAS~49  	   & 0.061  &  43.73 & $60_{-20}^{+60}$    & 1    & (r)   \\
NGC~4388	   & 0.0084 &  42.87 & $210_{-40}^{+120}$  & 1    & (r)   \\
NGC~4395	   & 0.0011 &  40.63 & $120_{-30}^{+50}$   & 1    & (r)   \\
LEDA~170194	   & 0.0367 &  43.2  & $<230$		   & 1    & (r)   \\
NGC~4941	   & 0.0037 &  41.73 & $<50$		   & 1    & (r)   \\
NGC~4992	   & 0.0251 &  43.46 & $80_{-30}^{+90}$    & 1    & (r)   \\
Mrk~248 	   & 0.0366 &  43.53 & $50_{-10}^{+20}$    & 1    & (r)   \\
ESO~509-IG066	   & 0.0446 &  43.63 & $70_{-20}^{+50}$    & 1    & (r)   \\
NGC~5252	   & 0.023  &  43.43 & $330_{-100}^{+150}$ & 1    & (r)   \\
2MASX~J1410	   & 0.0339 &  42.93 & $<80$		   & 1    & (r)   \\
NGC~5643	   & 0.004  &  41.13 & $<130$		   & 1    & (r)   \\
NGC~5674	   & 0.0249 &  43.23 & $<100$		   & 1    & (r)   \\
NGC~5728	   & 0.0094 &  43.23 & $80_{-20}^{+30}$    & 1    & (r)   \\
IC~4518A	   & 0.0163 &  42.73 & $120_{-50}^{+150}$  & 1    & (r)   \\
2MASX~J1506	   & 0.0377 &  42.93 & $<140$		   & 1    & (r)   \\
NGC~5899	   & 0.0086 &  42.23 & $<340$		   & 1    & (r)   \\
MCG+11-19-006	   & 0.044  &  43.43 & $<60$		   & 1    & (r)   \\
MCG-01-40-001	   & 0.0227 &  42.93 & $<130$		   & 1    & (r)   \\
NGC~5995	   & 0.0252 &  43.28 & $<340$		   & 1    & (r)   \\
MCG+14-08-004	   & 0.0239 &  42.81 & $<120$		   & 1    & (r)   \\
Mrk~1498	   & 0.0547 &  44.22 & $60_{-10}^{+10}$    & 1    & (r)   \\
IRAS~16288+3929    & 0.0306 &  43.37 & $<60$		   & 1    & (r)   \\
NGC~6240	   & 0.0245 &  44.02 & $90_{-30}^{+70}$    & 1    & (r)   \\
NGC~6300	   & 0.0037 &  42.04 & $210_{-50}^{+100}$  & 1    & (r)   \\
MCG+07-37-031	   & 0.0412 &  43.85 & $<110$		   & 1    & (r)   \\
2MASX~J1824	   & 0.067  &  43.83 & $<110$		   & 1    & (r)   \\
IC~4709 	   & 0.0169 &  42.83 & $140_{-60}^{+200}$  & 1    & (r)   \\
LEDA~3097193	   & 0.022  &  43.34 & $130_{-40}^{+110}$  & 1    & (r)   \\
ESO~103-G035	   & 0.0133 &  43.25 & $100_{-10}^{+20}$   & 1    & (r)   \\
ESO~231-G026	   & 0.0625 &  44.12 & $<250$		   & 1    & (r)   \\
2MASX~J1926	   & 0.071  &  43.33 & $<70$		   & 1    & (r)   \\
2MASX~J1947	   & 0.0539 &  43.83 & $120_{-40}^{+110}$  & 1    & (r)   \\
3C~403  	   & 0.059  &  44.03 & $<110$		   & 1    & (r)   \\
2MASX~J2006	   & 0.043  &  43.33 & $<80$		   & 1    & (r)   \\
2MASX~J2018	   & 0.0144 &  42.83 & $<100$		   & 1    & (r)   \\
2MASX~J2021	   & 0.017  &  42.63 & $<90$		   & 1    & (r)   \\
NGC~6921	   & 0.0145 &  43.13 & $<100$		   & 1    & (r)   \\
MCG+04-48-002	   & 0.0139 &  42.73 & $<150$		   & 1    & (r)   \\
IC~5063 	   & 0.0114 &  43.03 & $<130$		   & 1    & (r)   \\
NGC~7130	   & 0.0162 &  43.53 & $<100$		   & 1    & (r)   \\
MCG+06-49-019	   & 0.0213 &  42.13 & $<200$		   & 1    & (r)   \\
NGC~7319	   & 0.0225 &  42.63 & $<220$		   & 1    & (r)   \\
NGC~7582	   & 0.0053 &  41.78 & $200_{-80}^{+190}$  & 1    & (r)   \\
2MASX~J2330	   & 0.037  &  43.13 & $<70$		   & 1    & (r)   \\
PKS~2331-240	   & 0.0477 &  43.86 & $<110$		   & 1    & (r)   \\
PKS~2356-61	   & 0.0963 &  44.23 & $<80$		   & 1    & (r)   \\
   \end{longtable}
   \tablefoot{Col. (1): Source name; Col. (2): Redshift; Col. (3): Logarithm of the 2--10 keV luminosity ($\rm erg~s^{-1}$); Col. (4): High-energy cutoff (keV); Col. (5): telescope flag, 1={\it NuSTAR}, 2=nonfocusing; Col. (6): References. *: AGN with multiple observations in \citet{dadina2007}. Observation dates of selected measurement: 1996-12-06 for NGC~4151, 2000-11-08 for Mrk~509. }
 \tablebib{(j) \citet{dadina2007}, (k) \citet{kara2017}, (l) \citet{tortosa2018}, (m) \citet{middei2020}, (n) \citet{ursini2020}, (o) \citet{middei2021}, (p) \citet{reeves2021}, (q) \citet{kamraj2018}, (r) \citet{balokovic2020}.      }

\newpage

\begin{longtable}{lccccc}
\caption{High-$z$ measurements}
\label{tab:ecutplot3}\\
\hline\hline
Source & $z$ &  $\log L_{2-10}$  &  $E_{\rm cut}$  & Flag & References   \\
\hline
\endfirsthead

\caption*{continued.}\\
\hline\hline
Source & $z$ &  $\log L_{2-10}$  &  $E_{\rm cut}$  & Flag & References   \\
\hline
\endhead

\hline
\endfoot
2MASSJ16	   & 1.86   &  45.93 & $107_{-37}^{+102}$  & 1    & (b)    \\ 
B1422		   & 3.62   &  45.5  & $66_{-12}^{+17}$    & 1    & (b)    \\ 
APM~08279	   & 3.91   &  45.8  & $99_{-35}^{+91}$    & 1    & This work      \\
            \hline
   \end{longtable}
   \tablefoot{Col. (1): Source name; Col. (2): Redshift; Col. (3): Logarithm of the 2--10 keV luminosity ($\rm erg~s^{-1}$); Col. (4): High-energy cutoff (keV); Col. (5): telescope flag, 1={\it NuSTAR}, 2=nonfocusing; Col. (6): References. }
 \tablebib{(b) \citetalias{lanzuisi2019}. }

\end{appendix}

\end{document}